\documentclass[twocolumn,showpacs,preprintnumbers,amsmath,amssymb]{revtex4}

\usepackage{graphicx}
\usepackage{dcolumn}
\usepackage{bm}

\begin{document}

\preprint{APS/123-QED}

\title{Atomic spin squeezing in an optical cavity}

\author{Anne E. B. Nielsen and Klaus M{\o}lmer}
\affiliation{Lundbeck Foundation Theoretical Center for Quantum
System Research, Department of Physics and Astronomy, University of
Aarhus, DK-8000 \AA rhus C, Denmark}

\date{\today}

\begin{abstract}
We consider squeezing of one component of the collective spin vector
of an atomic ensemble inside an optical cavity. The atoms interact
with a cavity mode, and the squeezing is obtained by probing the
state of the light field that is transmitted through the cavity.
Starting from the stochastic master equation, we derive the time
evolution of the state of the atoms and the cavity field, and we
compute expectation values and variances of the atomic spin
components and the quadratures of the cavity mode. The performance
of the setup is compared to spin squeezing of atoms by probing of a
light field transmitted only once through the sample.
\end{abstract}


\maketitle

\section{Introduction}

Interactions between light and matter have several applications
within quantum information processing. When a light field interacts
with a collection of atoms, the light and the atoms become
entangled, and the state of the total system can no longer be
written as a direct product of quantum states of the individual
systems. As a consequence, if the light field is subsequently
subjected to measurements, the state of the atoms will also be
affected. This has, for instance, been utilized to entangle two
atomic ensembles \cite{duan,julsgaard} and to teleport the state of
a light field onto atoms \cite{teleportation}. It has also been
suggested to generate various squeezed and entangled states of light
and matter by sending a light field twice \cite{sherson2} or
multiple times \cite{hammerer} through the same atomic ensemble from
different directions.

\begin{figure}
\begin{center}
\includegraphics*[viewport=38 60 375 214,width=0.9\columnwidth]{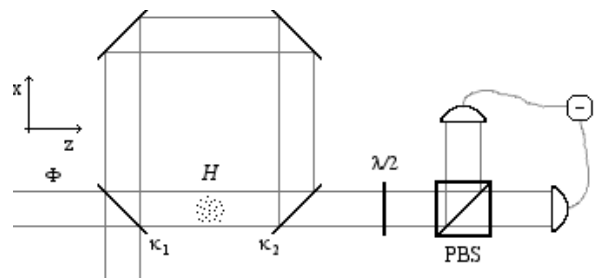}
\end{center}
\caption{Experimental setup for probing of the $z$-component of the
collective spin of an atomic ensemble with electromagnetic
radiation. A continuous laser beam linearly polarized in the
$x$-direction and with photon flux $\Phi$ enters the cavity from the
left. The interaction between the light and the atoms, described by
the Hamiltonian $H$, rotates the polarization vector of the light
field by an amount, which depends on the $z$-component of the atomic
spin. The angle of rotation can be measured by performing a
detection on the light leaking out of the cavity ($\kappa_1$ and
$\kappa_2$ denote cavity decay rates). The half wave plate
transforms the field operators $\hat{a}_x$ and $\hat{a}_y$ for $x$-
and $y$-polarized light into $(\hat{a}_x+\hat{a}_y)/\sqrt{2}$ and
$(\hat{a}_x-\hat{a}_y)/\sqrt{2}$, and these polarization components
are subsequently separated by the polarizing beam splitter (PBS).
The measurement outcome is the difference in photo current between
the two photo detectors.}\label{setup}
\end{figure}

The generation of entanglement between light and atoms may also be
utilized to perform a quantum nondemolition measurement of one of
the components of the collective spin vector of an atomic ensemble
\cite{kuzmich,takahashi,kuzmichexp,thomsen,geremia,hammerer,madsen,sherson}.
The measurements can reduce the uncertainty in the measured
observable below the uncertainty of a coherent spin state, resulting
in a squeezed spin state. Apart from the fundamental interest in
generating squeezed states, spin squeezing can improve the precision
of measurements of, for instance, weak magnetic fields
\cite{budker,partner}. The strength of the interaction between light
and atoms is normally weak but can be enhanced by placing the atoms
inside an optical cavity as depicted in Fig.\ \ref{setup}, and in
the present paper we investigate the performance of spin squeezing
in a cavity compared to spin squeezing in free space.

In spin squeezing experiments the usual initial state of the atoms
is a coherent spin state, where all the atomic spins are oriented in
the same direction, which we shall take as the $x$-direction. If the
number of atoms is large, the $x$-component of the collective atomic
spin $\hat{\boldsymbol{J}}=\sum_i\hat{\boldsymbol{j}}_i$, where
$\hat{\boldsymbol{j}}_i$ is the total spin of the $i$th atom, may be
treated as a classical quantity $\hat{J}_x\approx\langle
\hat{J}_x\rangle$, and the commutator between the scaled spin
components $\hat{x}_{\textrm{at}}=\hat{J}_y/(\hbar \langle
\hat{J}_x\rangle)^{1/2}$ and $\hat{p}_{\textrm{at}}=\hat{J}_z/(\hbar
\langle \hat{J}_x\rangle)^{1/2}$ turns into the canonical commutator
$[\hat{x}_{\textrm{at}},\hat{p}_{\textrm{at}}]=i$. In this
approximation the initial coherent spin state is a Gaussian state,
and since the interaction Hamiltonian and the measurements transform
Gaussian states into Gaussian states as long as $\hat{J}_x$ can be
treated classically, a very efficient Gaussian formalism, which
provides several analytical results for both pulsed and continuous
wave fields, is applicable, as demonstrated for free fields in
Refs.\ \cite{madsen,sherson}. The Gaussian description is easily
generalized to take an optical cavity into account \cite{nm4}, but
although we shall be mainly concerned with the limit of a large
number of atoms below, we also demonstrate that analytical results
can be obtained even without assuming the Gaussian approximation for
the collective atomic spin.

The paper is structured as follows. In Sec.\ \ref{II} we apply the
stochastic master equation for the setup in Fig.\ \ref{setup} to
derive expressions for the time evolution of the state of the atoms
and the cavity field, and in Sec.\ \ref{III} we evaluate the
variances and mean values of the collective atomic spin operators
and the quadratures of the cavity field as a function of time in the
limit of a large number of atoms. Effects of losses due to
spontaneous decay is considered in Sec.\ \ref{IV}, and the results
are compared to those obtained for squeezing in free space. Section
\ref{V} concludes the paper.

\section{Atoms interacting with off-resonant light in a cavity}\label{II}

We consider atoms with a spin $1/2$ ground state $|g_\mp\rangle$ and
a spin $1/2$ excited state $|e_\mp\rangle$ interacting with a
strong, off-resonant cavity field, which is initially linearly
polarized in the $x$-direction. Decomposing the light into right and
left circularly polarized cavity modes with field annihilation
operators $\hat{a}_+=(-\hat{a}_x+i\hat{a}_y)/\sqrt{2}$ and
$\hat{a}_-=(\hat{a}_x+i\hat{a}_y)/\sqrt{2}$, respectively, the
Hamiltonian takes the form
\begin{multline}\label{Hf}
H=\hbar g\sum_{i=1}^{N_{\textrm{at}}}
\left(\hat{a}_+|e_{+,i}\rangle\langle g_{-,i}|
+\hat{a}_-|e_{-,i}\rangle\langle g_{+,i}|+\textrm{h.c.}\right)\\
-\hbar\Delta\sum_{i=1}^{N_{\textrm{at}}}\left(|e_{+,i}\rangle\langle
e_{+,i}|+|e_{-,i}\rangle\langle e_{-,i}|\right)
\end{multline}
in a frame rotating with the frequency of the light field. The
summation runs over the $N_{\textrm{at}}$ atoms, $\hbar g=-dE_0$,
$d$ is the atomic dipole moment,
$E_0=\sqrt{\hbar\omega/(V\epsilon_0)}$, $\omega$ is the angular
frequency of the light field, $V$ is the mode volume, $\epsilon_0$
is the vacuum permittivity, and $\Delta=\omega-\omega_{\textrm{at}}$
is the detuning between the light field and the atomic transition.
For sufficiently large detuning $g/\Delta\ll1$, the excited states
will not become significantly populated and can be adiabatically
eliminated, which leads to the effective Hamiltonian
\begin{equation}\label{H+-}
H=\frac{\hbar g^2}{\Delta}\sum_{i=1}^{N_{\textrm{at}}}
\left(\hat{a}_+^\dag\hat{a}_+|g_{-,i}\rangle\langle g_{-,i}|
+\hat{a}_-^\dag\hat{a}_-|g_{+,i}\rangle\langle g_{+,i}|\right).
\end{equation}
In this and the next section, we neglect loss of photons and atomic
coherence due to spontaneous emission from the excited states, but
we return to an analysis of the role of spontaneous atomic emission
of light in Sec.\ \ref{IV}. Since $\sum_i|g_{\mp,i}\rangle\langle
g_{\mp,i}|=N_{\textrm{at}}/2\pm \hat{J}_z/\hbar$,
\begin{equation}\label{H}
H=\frac{\hbar g^2}{\Delta}\frac{N_{\textrm{at}}}{2}
(\hat{a}_x^\dag\hat{a}_x+\hat{a}_y^\dag\hat{a}_y)-i\frac{\hbar
g^2}{\Delta}
\left(\hat{a}_x^\dag\hat{a}_y-\hat{a}_y^\dag\hat{a}_x\right)
\frac{\hat{J}_z}{\hbar}.
\end{equation}
The first term gives rise to a common phase shift of the $x$- and
$y$-polarized light and can be compensated by introducing an
additional phase shift in the cavity, for instance by adjusting the
length of the cavity. We thus ignore this term in the following. For
the setup depicted in Fig.\ \ref{setup}, it is desirable to have a
large number of photons in the $x$-polarized mode, since this
increases the strength of the light-atom interaction, and since, in
the polarization rotation measurement, the $x$-polarized field acts
the same way as a local oscillator in balanced homodyne detection.
Approximating $\hat{a}_x$ by its expectation value
$\langle\hat{a}_x(t)\rangle$, the infinitesimal time evolution
operator corresponding to the second term in \eqref{H} takes the
form $\hat{U}=\exp\left((g^2/\Delta)
\left(\langle\hat{a}_x(t)\rangle\hat{a}_y^\dag-\langle\hat{a}_x^\dag(t)\rangle
\hat{a}_y\right)(\hat{J}_z/\hbar)dt\right)$, and comparing this to
the displacement operator
$\hat{D}(\delta)=\exp(\delta\hat{a}_y^\dag-\delta^*\hat{a}_y)$, we
observe that the interaction displaces the $y$-polarized field
amplitude by an amount, which is proportional to the $z$-component
of the atomic spin. Detecting the quadrature of the $y$-polarized
field in the direction of the displacement thus constitutes an
indirect measurement of $\hat{J}_z$ as stated above.

Assuming a high finesse cavity $\kappa\tau\ll1$, where $\kappa$ is
the total cavity decay rate and $\tau$ is the round trip time of
light in the cavity, and an only infinitesimal change of the atomic
quantum state on the time scale of $\tau$, we deduce from the
detailed derivation in \cite{nm6} that the density operator
$\rho(t)$, describing the state of the atoms and the $x$- and
$y$-polarized cavity fields, satisfies the linearized stochastic
master equation
\begin{multline}\label{SME}
d\rho(t)=-\frac{i}{\hbar}\left[H,\rho(t)\right]dt
+\sqrt{\eta_d\kappa_2}\left(\hat{a}_y\rho(t)dt+
\rho(t)\hat{a}_y^\dag\right)dy_s\\
+\frac{\kappa}{2}\left(-\hat{a}^\dag_y\hat{a}_y\rho(t)-
\rho(t)\hat{a}^\dag_y\hat{a}_y+2\hat{a}_y\rho(t)\hat{a}_y^\dag\right)dt\\
+\frac{\kappa}{2}\left(-\hat{a}^\dag_x\hat{a}_x\rho(t)-
\rho(t)\hat{a}^\dag_x\hat{a}_x+2\hat{a}_x\rho(t)\hat{a}_x^\dag\right)dt\\
+\sqrt{\kappa_1}\beta(t)[\hat{a}_x^\dag,\rho(t)]dt
-\sqrt{\kappa_1}\beta^*(t)[\hat{a}_x,\rho(t)]dt,
\end{multline}
where $\kappa=\kappa_1+\kappa_2+\kappa_L$, $\kappa_1$ ($\kappa_2$)
is the cavity decay rate due to the lower left (right) cavity mirror
in Fig.\ \ref{setup}, $\kappa_L$ is the cavity decay rate due to
additional losses, $\beta(t)$ is the amplitude of the incoming probe
beam, $\eta_d$ is the detector efficiency, and $dy_s$ is a
stochastic variable representing the measured difference in photo
current at time $t$ (see \cite{nm6} for details). The first term in
\eqref{SME} is the Hamiltonian evolution due to the interaction
between the atoms and the cavity modes, the second term represents
the knowledge obtained from the continuous measurement, the third
and fourth terms take cavity decay into account, and the fifth and
sixth terms appear due to the presence of the input beam. The
derivation in \cite{nm6} assumes that
$\sqrt{\kappa_1}|\beta(t)|\tau$ is small, but, for a classical
$x$-polarized mode, it is sufficient to assume that
$\sqrt{\kappa_1}\beta(t)$ varies slowly within times of order $\tau$
(and if we are not interested in features of the solution occurring
on a time scale $\tau$ or faster, we may even allow
$\sqrt{\kappa_1}\beta(t)$ to change abruptly). In fact, if the
$x$-polarized light is used as local oscillator as in Fig.\
\ref{setup}, it is required that
$4\kappa_1\kappa_2|\beta(t)|^2\tau/\kappa^2\gg1$, since the local
oscillator is assumed to be strong. We note that the stochastic term
in \eqref{SME} does not conserve the trace of the density operator,
which should hence be normalized explicitly. The probability to
obtain the normalized state $\rho(t+dt)/\textrm{Tr}(\rho(t+dt))$ at
time $t+dt$, given that the normalized state at time $t$ was
$\rho(t)$, is $\textrm{Tr}(\rho(t+dt))$ multiplied by the
probability to obtain the required value of $dy_s$, assuming that
$dy_s$ is a Gaussian distributed stochastic variable with zero mean
value and variance $dt$. If it is the reflected light and not the
transmitted light, which is subjected to measurement, and if the
lower right cavity mirror in Fig.\ \ref{setup} is perfectly
reflecting, we note that $\eta_d\kappa_2$ should be replaced by
$\eta_d\kappa_1$ and $\kappa=\kappa_1+\kappa_2+\kappa_L$ should be
replaced by $\kappa=\kappa_1+\kappa_L$ in Eq.\ \eqref{SME}.

Since the Hamiltonian \eqref{H} commutes with
$\hat{\boldsymbol{J}}^2$, we can restrict ourselves to the basis
consisting of the states with total spin quantum number
$J=N_{\textrm{at}}/2$ if the initial state is a coherent spin state.
We thus consider the states $J_z|n\rangle=\hbar n|n\rangle$ with
$n=-N_{\textrm{at}}/2,-N_{\textrm{at}}/2+1,\ldots,N_{\textrm{at}}/2$,
and write the density matrix as
\begin{equation}\label{dec}
\rho(t)=\sum_n\sum_m\rho_{nm}|n\rangle\langle m|,
\end{equation}
where $\rho_{nm}$ are operators on the space of the $x$- and
$y$-polarized cavity field modes. This leads to the
$(N_{\textrm{at}}+1)^2$ independent equations
\begin{multline}
d\rho_{nm}=-\frac{g^2}{\Delta}n(\hat{a}_x^\dag\hat{a}_y-
\hat{a}_y^\dag\hat{a}_x)\rho_{nm}dt\\
+\frac{g^2}{\Delta}m\rho_{nm}(\hat{a}_x^\dag\hat{a}_y-\hat{a}_y^\dag\hat{a}_x)dt\\
+\sqrt{\eta_d\kappa_2}(\hat{a}_y\rho_{nm}+\rho_{nm}\hat{a}_y^\dag)dy_s\\
+\frac{\kappa}{2}\left(-\hat{a}_y^\dag\hat{a}_y\rho_{nm}
-\rho_{nm}\hat{a}^\dag_y\hat{a}_y+2\hat{a}_y\rho_{nm}\hat{a}_y^\dag\right)dt\\
+\frac{\kappa}{2}\left(-\hat{a}_x^\dag\hat{a}_x\rho_{nm}
-\rho_{nm}\hat{a}^\dag_x\hat{a}_x+2\hat{a}_x\rho_{nm}\hat{a}_x^\dag\right)dt\\
+\sqrt{\kappa_1}\beta(t)[\hat{a}_x^\dag,\rho_{nm}]dt
-\sqrt{\kappa_1}\beta^*(t)[\hat{a}_x,\rho_{nm}]dt
\end{multline}
with solution
\begin{equation}\label{rhonm}
\rho_{nm}=C_{nm}(t)|\gamma_n(t)\rangle_x\langle\gamma_m(t)|
\otimes|\alpha_n(t)\rangle_y\langle\alpha_m(t)|,
\end{equation}
where $|\gamma_n(t)\rangle$ and $|\alpha_n(t)\rangle$ are coherent
states satisfying
\begin{equation}\label{gamman}
\frac{d\gamma_n(t)}{dt}=-\frac{\kappa}{2}\gamma_n(t)-n\frac{g^2}{\Delta}
\alpha_n(t)+\sqrt{\kappa_1}\beta(t)
\end{equation}
and
\begin{equation}
\frac{d\alpha_n(t)}{dt}=-\frac{\kappa}{2}\alpha_n(t)+n\frac{g^2}{\Delta}
\gamma_n(t).
\end{equation}
For a classical input field the term in \eqref{gamman} proportional
to $\alpha_n(t)$ is negligible, and, assuming
$\beta(t)=\beta^*(t)=\sqrt{\Phi(t)}$ and
$\alpha_n(0)=\gamma_n(0)=0$, we obtain
\begin{equation}
\langle\hat{a}_x(t)\rangle=\gamma_n(t)
=\sqrt{\kappa_1}\int_0^te^{-\kappa(t-t')/2}\beta(t')dt'
\end{equation}
and
\begin{equation}\label{alfan}
\alpha_n(t)=n\frac{g^2}{\Delta}\int_0^te^{-\kappa(t-t')/2}
\langle\hat{a}_x(t')\rangle dt'\equiv n\alpha(t),
\end{equation}
where $\alpha(t)$ is real and independent of $n$. Under these
conditions the coefficients $C_{nm}$ in \eqref{rhonm} satisfy
\begin{equation}
\frac{dC_{nm}}{C_{nm}}= \sqrt{\eta_d\kappa_2}(n+m)\alpha(t)dy_s
-\frac{\kappa}{2}(n-m)^2\alpha(t)^2dt
\end{equation}
with solution
\begin{multline}\label{cnm}
C_{nm}(t)=C_{nm}(0)\exp\bigg(-\frac{\kappa}{2}(n-m)^2\int_0^t\alpha(t')^2dt'\\
+\sqrt{\eta_d\kappa_2}(n+m)\int_0^t\alpha(t')dy_s'\\
-\frac{\eta_d\kappa_2}{2}(n+m)^2\int_0^t\alpha(t')^2dt'\bigg).
\end{multline}
If required, the analysis is easily generalized to include all
simultaneous eigenstates of $\hat{\boldsymbol{J}}^2$ and
$\hat{J}_z$, since it is only needed to include more terms in
\eqref{dec} and to introduce additional labels to distinguish the
different states.

We finally note that if $\Phi(t)$ is zero for $t<0$ and assumes the
constant value $\Phi$ for $t>0$ and if the light-atom coupling is
sufficiently weak to ensure that the change in the state of the
atoms during the transient is negligible, we may approximate
$\langle\hat{a}_x(t)\rangle$ and $\alpha(t)$ by their respective
steady state values
$\langle\hat{a}_x\rangle=2\sqrt{\kappa_1\Phi}/\kappa$ and
$\alpha=2g^2\langle\hat{a}_x\rangle/(\kappa\Delta)$ for $t>0$, which
makes the integrals in \eqref{cnm} trivial to evaluate. In that case
the state at time $t$ depends on the measurement result through the
integrated signal $Y_s=\int_0^tdy_s$ only, and the probability
density to measure a given value of $Y_s$ is \cite{nm6}
\begin{equation}\label{P}
P=\sum_n\frac{C_{nn}(0)}{\sqrt{2\pi t}} \exp\left(-\frac{(Y_s-
2\sqrt{\eta_d\kappa_2}n\alpha t)^2}{2t}\right).
\end{equation}
The measurement leads to a narrowing of the distribution
$C_{nn}(t)/\sum_mC_{mm}(t)$ over the possible eigenstates of
$\hat{J}_z$, but the expectation value of $\hat{J}_z$ depends on
$Y_s$, and if we average over all possible measurement outcomes, we
find that $C_{nn}(t)=C_{nn}(0)$.

\section{Expectation values of atomic spin operators
for large $N_{\textrm{at}}$}\label{III}

Having obtained the state of the atoms and the $y$-polarized cavity
field as a function of time, we can now evaluate expectation values
and variances of the atomic spin operators and the field quadrature
operators
$\hat{x}_{\textrm{ph}}=(\hat{a}_y+\hat{a}_y^\dag)/\sqrt{2}$ and
$\hat{p}_{\textrm{ph}}=-i(\hat{a}_y-\hat{a}_y^\dag)/\sqrt{2}$. We
assume below that the initial state is a coherent spin state
pointing in the $x$-direction and that the number of atoms is large
$N_{\textrm{at}}\gg1$, since this is a typical experimental
condition, and since it allows us to simplify the obtained
expressions considerably. In order to stay within the parameter
regime where the Gaussian approximation, discussed in the
Introduction, is valid, it is also required that the total
measurement time is short compared to the time it takes to gain
sufficient information to project the state of the atoms onto a
single eigenstate of $\hat{J}_z$. For the steady state case it
follows from Eq.\ \eqref{P} that the relevant time scale is
determined by the condition $4\eta_d\kappa_2\alpha^2t\sim1$, and we
thus assume in the following that
$4\eta_d\kappa_2\int_0^t\alpha(t')^2dt'$ is small, i.e., comparable
to the size of $N_{\textrm{at}}^{-1}$, while
$2\sqrt{\eta_d\kappa_2}\int_0^t\alpha(t')d\hat{y}_s'$ is assumed to
be comparable to $N_{\textrm{at}}^{-1/2}$.

First we would like to determine whether the atomic spin is indeed
squeezed, and we thus trace out the cavity field and compute the
variance of $\hat{J}_z$
\begin{equation}\label{varJzn}
\textrm{Var}\left(\hat{J}_z/\hbar\right)=
\frac{\sum_nn^2C_{nn}(t)}{\sum_nC_{nn}(t)}
-\left(\frac{\sum_nnC_{nn}(t)}{\sum_nC_{nn}(t)}\right)^2.
\end{equation}
For a coherent spin state pointing in the $x$-direction
\begin{multline}
C_{nm}(0)=\frac{1}{2^{N_{\textrm{at}}}}\sqrt{\frac{N_{\textrm{at}}!}
{(N_{\textrm{at}}/2+n)!(N_{\textrm{at}}/2-n)!}}\\
\times\sqrt{\frac{N_{\textrm{at}}!}
{(N_{\textrm{at}}/2+m)!(N_{\textrm{at}}/2-m)!}},
\end{multline}
we may apply the approximation
\begin{equation}\label{cnm0}
C_{nm}(0)\approx\sqrt{\frac{2}{\pi
N_{\textrm{at}}}}\exp\left(-\frac{n^2+m^2}{N_{at}}\right),
\end{equation}
and it follows from \eqref{cnm}, \eqref{varJzn}, and \eqref{cnm0}
that
\begin{equation}\label{varJz}
\frac{\textrm{Var}\left(\hat{J}_z/\hbar\right)}{N_{\textrm{at}}/2}=
\frac{1}{2}\left(1+N_{\textrm{at}}\eta_d\kappa_2
\int_0^t\alpha(t')^2dt'\right)^{-1}.
\end{equation}
Remarkably, this result does not depend on the measurement readout
and is thus deterministic. The variance of
$\hat{J}_z/\hbar/\sqrt{N_{\textrm{at}}/2}$ is seen to be decreasing
and smaller than $1/2$ at all times $t>0$ if $\eta_d>0$. For the
special case where $\Phi(t)$ is zero for $t<0$ and assumes the
constant value $\Phi$ for $t>0$, we have
$\langle\hat{a}_x(t)\rangle=\sqrt{4\kappa_1\Phi/\kappa^2}(1-\exp(-\kappa
t/2))$, and
\begin{equation}
\frac{\textrm{Var}\left(\hat{J}_z/\hbar\right)}{N_{\textrm{at}}/2}=
\frac{1}{2}\left(1+N_{\textrm{at}}\frac{4\kappa_1\Phi}{\kappa^2}
\frac{4g^4}{\kappa^2\Delta^2}\frac{\eta_d\kappa_2}{\kappa}\kappa
\tilde{t}\right)^{-1},
\end{equation}
where
\begin{equation}
\tilde{t}\equiv t-\frac{11}{2\kappa}+\frac{2\kappa
t+8}{\kappa}e^{-\kappa t/2}-\frac{\kappa^2t^2+6\kappa
t+10}{4\kappa}e^{-\kappa t}.
\end{equation}
This is to be compared to the expression
\begin{equation}
\left(\frac{\textrm{Var}(\hat{J}_z/\hbar)}{N_{\textrm{at}}/2}\right)_{\textrm{sp}}=
\frac{1}{2}\left(1+N_{\textrm{at}}\Phi\frac{g^4\tau^2}{\Delta^2}\eta_d
t\right)^{-1}
\end{equation}
for single-pass squeezing \cite{madsen}. Apart from what effectively
amounts to a small reduction of the probing time, appearing because
it takes a short while to build up the cavity field, the effect of
the cavity is to increase the coefficient multiplying $t$ by a
factor $Q=16\kappa_1\kappa_2/(\kappa^4\tau^2)$. In the single-pass
case each segment of temporal width $\tau$ of the probe beam
interacts only once with the atoms, and
$\langle\hat{a}_x\rangle=\sqrt{\Phi\tau}$ for all times $t>0$. The
interaction thus transforms the $y$-polarized mode from the vacuum
state $|0\rangle$ into the coherent state
$\hat{U}|0\rangle=|n(g^2\tau/\Delta)\sqrt{\Phi\tau}\rangle$, where,
for simplicity, we have assumed that the atoms are in the
$\hat{J}_z$ eigenstate $|n\rangle$. The number of $y$-polarized
photons observed per unit time is thus
$n^2(g^4\tau^2/\Delta^2)\Phi\eta_d$. If the cavity is included, on
the other hand, the number of $y$-polarized photons observed per
unit time is the product of the number of $y$-polarized photons in
the cavity $|\alpha_n|^2$, the rate $\kappa_2$ with which the
photons leave the cavity through the cavity output mirror in Fig.\
\ref{setup}, and the detector efficiency $\eta_d$, and the result
$n^2(4g^4/(\kappa^2\Delta^2))(4\kappa_1\Phi/\kappa^2)\eta_d\kappa_2$
is larger than in the single-pass case by precisely the factor $Q$.
To understand this increase in the number of detected $y$-polarized
photons, $Q$ may be divided into the three factors
$\kappa_2/\kappa$, $4\kappa_1/(\kappa^2\tau)$, and $4/(\kappa\tau)$,
where the first appears because the effective detector efficiency is
$\eta_d\kappa_2/\kappa$ for squeezing in a cavity and $\eta_d$ for
single-pass squeezing, the second factor appears due to the increase
in the number of photons in the $x$-polarized mode, as can be seen
from the increase in production rate of $y$-polarized photons, when
the flux of $x$-polarized photons in the case of single-pass
squeezing is increased from $\Phi$ to
$4\kappa_1\Phi/(\kappa^2\tau)$, and the third factor appears because
photons are present in the $y$-polarized mode in the cavity, as can
be seen by comparing the number of produced $y$-polarized photons
when $\hat{U}$ acts on $|\alpha_n\rangle$ and when $\hat{U}$ acts on
$|0\rangle$.

Applying $C_{n\pm1n\mp1}(0)=
C_{nn}(0)(1-2/N_{\textrm{at}}+O(N_{\textrm{at}}^{-2}))$, we also
find
\begin{equation}\label{varJy}
\frac{\textrm{Var}\left(\hat{J}_y/\hbar\right)}{N_{\textrm{at}}/2}=
\frac{1}{2}\left(1+N_{\textrm{at}}\kappa
\int_0^t\alpha(t')^2dt'+N_{\textrm{at}}\alpha(t)^2\right).
\end{equation}
Since $\kappa$ is larger than or equal to $\eta_d\kappa_2$, the
product of \eqref{varJz} and \eqref{varJy} is larger than or equal
to $1/4$ as required by the Heisenberg uncertainty relation.
Equality is only obtained for $\alpha(t)=0$ and
$\kappa=\eta_d\kappa_2$, where the first equation is satisfied if
the $y$-polarized cavity mode is in the vacuum state at the final
time $t$, and the second equation is satisfied if all photons that
leave the cavity are detected.

It follows from
\begin{equation}
\textrm{Var}(\hat{x}_{\textrm{ph}})=\frac{1}{2}
\frac{1+N_{\textrm{at}}\alpha(t)^2+
N_{\textrm{at}}\eta_d\kappa_2\int_0^t\alpha(t')^2dt'}
{1+N_{\textrm{at}}\eta_d\kappa_2\int_0^t\alpha(t')^2dt'}
\end{equation}
and
\begin{equation}
\textrm{Var}(\hat{p}_{\textrm{ph}})=1/2
\end{equation}
that the cavity field is not squeezed, but, for time independent
$\alpha(t)$, the uncertainty in $\hat{x}_\textrm{ph}$ decreases with
probing time. The Heisenberg limit is only achieved exactly at
times, where the cavity field is in the vacuum state.

The expectation values
\begin{eqnarray}
\langle\hat{J}_y/\hbar\rangle&=&0\\
\langle\hat{J}_z/\hbar\rangle&=&
\frac{\sqrt{\eta_d\kappa_2}\int_0^t\alpha(t')dy_s'}
{2/N_{\textrm{at}}+2\eta_d\kappa_2\int_0^t\alpha(t')^2dt'}\\
\langle\hat{x}_{\textrm{ph}}\rangle&=&\sqrt{2}\alpha(t)
\langle\hat{J}_z/\hbar\rangle\\
\langle\hat{p}_{\textrm{ph}}\rangle&=&0
\end{eqnarray}
are either stochastic or zero, depending on whether the measurements
supply information on the concerned operator or not. Different mean
values of $J_z$ are thus obtained if the experiment is repeated. By
applying feedback and rotating the collective spin, it is, however,
possible to achieve absolute squeezing, where the same mean value of
$J_z$ is obtained in each run \cite{thomsen,geremia}.

An alternative approach to calculate expectation values and
variances of $\hat{x}_{\textrm{at}}=\hat{J}_y/(\hbar \langle
\hat{J}_x\rangle)^{1/2}$, $\hat{p}_{\textrm{at}}=\hat{J}_z/(\hbar
\langle \hat{J}_x\rangle)^{1/2}$, $\hat{x}_{\textrm{ph}}$, and
$\hat{p}_{\textrm{ph}}$ is to assume from the start that the state
of the atoms and the $y$-polarized cavity mode is approximately
Gaussian at all times $t$ satisfying
$4\eta_d\kappa_2\int_0^t\alpha(t')^2dt'\ll1$. Gaussian states are
efficiently described in terms of Wigner functions, and we thus
translate the nonlinear stochastic master equation for the density
operator for the atoms and the $y$-polarized cavity mode derived in
\cite{nm6}
\begin{multline}\label{NLSME}
d\rho(t)=-\frac{i}{\hbar}\left[H,\rho(t)\right]dt\\
+\sqrt{\eta_d\kappa_2}\left(\hat{a}\rho(t)-\mathrm{Tr}
\left(\hat{a}\rho(t)\right)\rho(t)\right)dW_s\\
+\sqrt{\eta_d\kappa_2} \left(\rho(t)\hat{a}^\dag-\mathrm{Tr}
\left(\rho(t)\hat{a}^\dag\right)\rho(t)\right)dW_s\\
+\frac{1}{2}\kappa\left(-\hat{a}^\dag\hat{a}\rho(t)-
\rho(t)\hat{a}^\dag\hat{a}+2\hat{a}\rho(t)\hat{a}^\dag\right)dt,
\end{multline}
where $dW_s$ is a Gaussian distributed stochastic variable with zero
mean value and variance $dt$, into an equation involving the Wigner
function $W$
\begin{multline}\label{dWdt}
dW=-\tilde{g}(t)\left(p_{\textrm{at}}\frac{\partial}{\partial
x_{\textrm{ph}}}+p_{\textrm{ph}}
\frac{\partial}{\partial x_{\textrm{at}}}\right)Wdt\\
+\sqrt{2\eta_d\kappa_2}\left(x_{\textrm{ph}}-\langle
\hat{x}_{\textrm{ph}}\rangle+\frac{1}{2}\frac{\partial}{\partial
x_{\textrm{ph}}}\right)WdW_s\\
+\kappa\Bigg(1+\frac{1}{2}\left(x_{\textrm{ph}}\frac{\partial}{\partial
x_{\textrm{ph}}}+p_{\textrm{ph}}\frac{\partial}{\partial p_{\textrm{ph}}}\right)\\
+\frac{1}{4}\left(\frac{\partial^2}{\partial x_{\textrm{ph}}^2}
+\frac{\partial^2}{\partial p_{\textrm{ph}}^2}\right)\Bigg)Wdt,
\end{multline}
where $W$ is a function of $t$ and the quadrature variables
$x_{\textrm{at}}$, $p_{\textrm{at}}$, $x_{\textrm{ph}}$, and
$p_{\textrm{ph}}$, and we have introduced the effective light atom
coupling strength
\begin{equation}\label{gtilde}
\tilde{g}(t)=\frac{2g^2}{\Delta}\sqrt{\frac{\langle
J_x\rangle}{\hbar}}\frac{\langle\hat{a}_x(t)\rangle}{\sqrt{2}},
\end{equation}
in terms of which the Hamiltonian reads
$H=\hbar\tilde{g}(t)\hat{p}_{\textrm{at}}\hat{p}_{\textrm{ph}}$. For
a Gaussian state
\begin{equation}\label{W}
W=\frac{1}{\pi^2\sqrt{\det(V)}}
\exp(-(y-\langle\hat{y}\rangle)^TV^{-1}(y-\langle\hat{y}\rangle)),
\end{equation}
where
$y=(x_{\textrm{at}},p_{\textrm{at}},x_{\textrm{ph}},p_{\textrm{ph}})^T$
is a column vector of quadrature variables,
$\hat{y}=(\hat{x}_{\textrm{at}},\hat{p}_{\textrm{at}},
\hat{x}_{\textrm{ph}},\hat{p}_{\textrm{ph}})^T$ is a column vector
of the corresponding quadrature operators, and
$V=\langle(\hat{y}-\langle\hat{y}\rangle)(\hat{y}-\langle\hat{y}\rangle)^T\rangle
+\langle(\hat{y}-\langle\hat{y}\rangle)(\hat{y}-\langle\hat{y}\rangle)^T\rangle^T$
is the covariance matrix. Inserting \eqref{W} into \eqref{dWdt}, we
find that
\begin{equation}\label{Ricatti}
\frac{dV}{dt}=G-DV-VE-VFV,
\end{equation}
where
\begin{equation}
G=\left[\begin{array}{cccc}0&0&0&0\\0&0&0&0\\0&0&\kappa-\eta\kappa_2&0\\
0&0&0&\kappa\end{array}\right],
\end{equation}
\begin{equation}
D=E^T=\left[\begin{array}{cccc}0&0&0&-\tilde{g}(t)\\0&0&0&0\\0&-\tilde{g}(t)
&\kappa/2-\eta_d\kappa_2&0\\
0&0&0&\kappa/2\end{array}\right],
\end{equation}
\begin{equation}\label{F}
F=\left[\begin{array}{cccc}0&0&0&0\\0&0&0&0\\0&0&\eta_d\kappa_2&0\\
0&0&0&0\end{array}\right],
\end{equation}
and
\begin{multline}
d\langle\hat{y}\rangle=
\left[\begin{array}{cccc}0&0&0&\tilde{g}(t)\\0&0&0&0\\0&\tilde{g}(t)&-\kappa/2&0\\
0&0&0&-\kappa/2\end{array}\right]\langle\hat{y}\rangle dt\\
+\sqrt{\frac{\eta_d\kappa_2}{2}}(V-I)
\left[\begin{array}{c}0\\0\\1\\0\end{array}\right]dW_s.
\end{multline}
Equation \eqref{Ricatti} is a so-called matrix Ricatti equation, and
if $V$ is decomposed according to $V=MK^{-1}$, it can be rewritten
as the linear set of equations $\dot{M}=-DM+GK$ and $\dot{K}=FM+EK$.
Solving these equations analytically for a time independent
$\tilde{g}$, we find expressions, which are in accordance with the
above results. Equation \eqref{Ricatti} can also be derived
following the covariance matrix approach outlined in Ref.\
\cite{madsen} for single-pass interaction. To do so, the light beams
are divided into segments of duration $\tau$, where each segment
constitutes a classical $x$-polarized field mode and a quantum
mechanical $y$-polarized field mode, and the state of the atomic
spin and the quantum mechanical field modes is assumed to be
Gaussian. The time evolution of the covariance matrix is then
obtained by realizing that an interaction between the atoms and the
field modes, a beam splitter operation, and a homodyne detection of
a field mode all amount to certain transformations of the covariance
matrix.

\section{Inclusion of loss due to spontaneous decay}\label{IV}

If the atoms are allowed to decay by spontaneous emission, there
will be a loss of atomic coherence as well as a decay of the mean
spin, because the polarization of a spontaneously emitted photon, in
principle, provides information on the final state of the atom that
emitted the photon. To include spontaneous emission in the analysis,
we add decay terms to the master equation for interaction of the
atoms with an $x$-polarized and a $y$-polarized light mode
\begin{multline}\label{decay}
\frac{d\rho}{dt}=-\frac{i}{\hbar}[H,\rho]\\
+\Gamma\bigg(\frac{2}{3}D[|g_{-,i}\rangle\langle e_{+,i}|]\rho
+\frac{1}{3}D[|g_{+,i}\rangle\langle e_{+,i}|]\rho\\
+\frac{2}{3}D[|g_{+,i}\rangle\langle e_{-,i}|]\rho
+\frac{1}{3}D[|g_{-,i}\rangle\langle e_{-,i}|]\rho\bigg),
\end{multline}
where $H$ is given by \eqref{Hf} and
\begin{equation}
D[\hat{c}]\rho\equiv\hat{c}\rho\hat{c}^\dag-
(\hat{c}^\dag\hat{c}\rho+\rho\hat{c}^\dag\hat{c})/2.
\end{equation}
Adiabatic elimination of the exited atomic states leads to
\begin{multline}\label{MEWL}
d\rho=-\frac{g^2\Delta}{\Delta^2+\frac{\Gamma^2}{4}}
\left[\left(\hat{a}_x^\dag\hat{a}_y-\hat{a}_y^\dag\hat{a}_x\right)
\frac{\hat{J}_z}{\hbar},\rho\right]dt
+\frac{\Gamma}{2}\frac{g^2}{\Delta^2+\frac{\Gamma^2}{4}}\\
\times\Bigg(-\hat{a}_-^\dag\hat{a}_-\left(\frac{N_{at}}{2}
-\frac{\hat{J}_z}{\hbar}\right)\rho-\rho\left(\frac{N_{at}}{2}
-\frac{\hat{J}_z}{\hbar}\right)\hat{a}_-^\dag\hat{a}_-\\
-\hat{a}_+^\dag\hat{a}_+\left(\frac{N_{at}}{2}
+\frac{\hat{J}_z}{\hbar}\right)\rho-\rho\left(\frac{N_{at}}{2}
+\frac{\hat{J}_z}{\hbar}\right)\hat{a}_+^\dag\hat{a}_+\Bigg)dt\\
+\frac{\Gamma}{2}\frac{g^2}{\Delta^2+\frac{\Gamma^2}{4}}\sum_{i=1}^{N_{at}}\bigg(
\frac{4}{3}\hat{a}_-|g_{+,i}\rangle\langle g_{+,i}|\rho
|g_{+,i}\rangle\langle g_{+,i}|\hat{a}_-^\dag\\
+\frac{4}{3}\hat{a}_+|g_{-,i}\rangle\langle g_{-,i}|\rho
|g_{-,i}\rangle\langle g_{-,i}|\hat{a}_+^\dag\\
+\frac{2}{3}\hat{a}_-|g_{-,i}\rangle\langle g_{+,i}|\rho
|g_{+,i}\rangle\langle g_{-,i}|\hat{a}_-^\dag\\
+\frac{2}{3}\hat{a}_+|g_{+,i}\rangle\langle g_{-,i}|\rho
|g_{-,i}\rangle\langle g_{+,i}|\hat{a}_+^\dag\bigg)dt,
\end{multline}
where, as before, $g/\Delta\ll1$ and we have omitted the term in the
Hamiltonian giving rise to a common phase shift of the light modes.
Finally, homodyne detection, cavity decay, and the input beam are
taken into account by adding the terms
\begin{align}\label{et}
&\sqrt{\eta_d\kappa_2}\left(\hat{a}\rho(t)-\mathrm{Tr}
\left(\hat{a}\rho(t)\right)\rho(t)\right)dW_s\nonumber\\
+&\sqrt{\eta_d\kappa_2} \left(\rho(t)\hat{a}^\dag-\mathrm{Tr}
\left(\rho(t)\hat{a}^\dag\right)\rho(t)\right)dW_s\nonumber\\
+&\frac{\kappa}{2}\left(-\hat{a}^\dag_y\hat{a}_y\rho(t)-
\rho(t)\hat{a}^\dag_y\hat{a}_y+2\hat{a}_y\rho(t)\hat{a}_y^\dag\right)dt\nonumber\\
+&\frac{\kappa}{2}\left(-\hat{a}^\dag_x\hat{a}_x\rho(t)-
\rho(t)\hat{a}^\dag_x\hat{a}_x+2\hat{a}_x\rho(t)\hat{a}_x^\dag\right)dt\nonumber\\
+&\sqrt{\kappa_1}\beta(t)[\hat{a}_x^\dag,\rho(t)]dt
-\sqrt{\kappa_1}\beta^*(t)[\hat{a}_x,\rho(t)]dt
\end{align}
on the right hand side of \eqref{MEWL}.

\begin{figure}
\begin{center}
\includegraphics*[viewport=15 2 398 320,width=0.95\columnwidth]{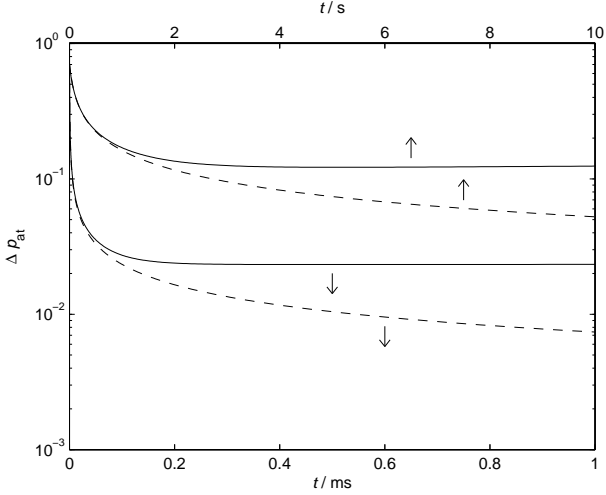}
\end{center}
\caption{Uncertainty in $p_{\textrm{at}}$ as a function of time with
atomic decay included (solid curves) and excluded (dashed curves).
The upper and lower curves represent squeezing of the same atomic
system in free space and in a cavity, respectively. Note the
different time scales. The parameters are (see \cite{madsen}):
$N_{at}=10^{12}$, $\Phi=10^{14}\textrm{ s}^{-1}$ for $t>0$,
$A=2\textrm{ cm}^2$, $\tau=3\cdot10^{-10}\textrm{ s}$,
$\Delta=2\pi\cdot10^{10}\textrm{ Hz}$, $\lambda=852\textrm{ nm}$,
$\Gamma=3.1\cdot10^7\textrm{ s}^{-1}$, $d=2.61\cdot10^{-29}\textrm{
Cm}$, $\kappa=\kappa_1=2\pi\cdot3\cdot10^6\textrm{ Hz}$ (i.e., we
observe the reflected light and assume $\kappa_L=0$), and
$\eta_d=1$.}\label{sqpmkavitet}
\end{figure}

Equation \eqref{MEWL}, \eqref{et} can be solved numerically for a
small number of atoms and a classical $x$-polarized mode, but here
we aim at an approximate description, which is valid for the case,
where the $x$-polarized mode is classical, the initial atomic state
is a coherent spin state pointing in the $x$-direction, $N_{at}$ is
sufficiently large to assume that $\hat{J}_x$ is classical, and $t$
is small compared to the time it takes to project the atomic state
onto an eigenstate of $\hat{J}_z$ due to measurements and small
compared to the time it takes $\langle\hat{J}_x\rangle$ to decay
significantly. From the stochastic mater equation it follows that
\begin{multline}\label{Jx}
\frac{1}{\hbar}\frac{d\langle\hat{J}_x(t)\rangle}{dt}=i\langle\hat{a}_x(t)\rangle
\frac{g^2\Delta}{\Delta^2+\Gamma^2/4}
\frac{\langle(\hat{a}_y-\hat{a}_y^\dag)\hat{J}_y\rangle}{\hbar}\\
-\langle\hat{a}_x(t)\rangle^2\frac{\Gamma}{2}\frac{g^2}{\Delta^2+\Gamma^2/4}
\frac{\langle\hat{J}_x(t)\rangle}{\hbar}.
\end{multline}
The ratio between the last and the first term is approximately
$\langle\hat{a}_x(t)\rangle\langle\hat{J}_x(t)\rangle^{1/2}\Gamma/(2\Delta)$,
which evaluates to $10^6$ for the parameters given in the caption of
Fig.\ \ref{sqpmkavitet}, and we thus skip the first term and obtain
\begin{equation}
\langle\hat{J}_x(t)\rangle=\frac{\hbar
N_{\textrm{at}}}{2}\exp\left(-\int_0^t\eta(t')dt'\right),
\end{equation}
where we have defined the time dependent decay rate $\eta(t)$ of the
atomic spin as
\begin{equation}\label{eta}
\eta(t)=\langle\hat{a}_x(t)\rangle^2\frac{\Gamma}{2}\frac{g^2}{\Delta^2+\Gamma^2/4}.
\end{equation}
Similarly, for $\langle\hat{a}_x(t)\rangle$ we find
\begin{equation}
\frac{d\langle\hat{a}_x(t)\rangle}{dt}=
-\frac{\kappa+\epsilon}{2}\langle\hat{a}_x(t)\rangle
+\sqrt{\kappa_1\Phi},
\end{equation}
where we have defined the photon absorption rate as
\begin{equation}
\epsilon=N_{\textrm{at}}\frac{\Gamma}{2}\frac{g^2}{\Delta^2+\Gamma^2/4}.
\end{equation}
We can now use the stochastic master equation to derive expressions
for the time derivative of the first and second order moments of
$\hat{J}_y$, $\hat{J}_z$, $\hat{x}_{\textrm{ph}}$, and
$\hat{p}_{\textrm{ph}}$, and we find that, apart from third order
moments appearing in the stochastic terms of the equations for the
time derivative of the second order moments, these expressions
contain only first and second order moments. Since the state of the
atoms and the light field is nearly Gaussian under the above
conditions, we approximate the third order moments by a sum over
products of first and second order moments to obtain a closed set of
equations. We also approximate $V_{12}$, $V_{13}$, $V_{24}$, and
$V_{34}$ by zero, because these covariance matrix elements are zero
if spontaneous emission is neglected, and because the rest of the
covariance matrix elements only couple to $V_{12}$, $V_{13}$,
$V_{24}$, and $V_{34}$ through terms that are proportional to the
small factor $(\Gamma/2)g^2/(\Delta^2+\Gamma^2/4)$. Within these
approximations we find that the time evolution of the covariance
matrix is given by the Ricatti equation \eqref{Ricatti} with
\begin{multline}
G=\frac{\hbar N_{\textrm{at}}}{\langle\hat{J}_x(t)\rangle}
\left[\begin{array}{cccc}\eta(t)&0&0&0\\
0&\frac{2}{3}\eta(t)&0&0\\0&0&0&0\\
0&0&0&0\end{array}\right]\\
+\left[\begin{array}{cccc}0&0&0&0\\
0&0&0&0\\0&0&\kappa+\epsilon-\eta_d\kappa_2&0\\
0&0&0&\kappa+\epsilon\end{array}\right],
\end{multline}
\begin{multline}
D=E^T=\\
\left[\begin{array}{cccc}\eta(t)/2&0&0&-\tilde{g}(t)\\
0&\eta(t)/6&0&0\\0&-\tilde{g}(t)&(\kappa+\epsilon)/2-\eta_d\kappa_2&0\\
0&0&0&(\kappa+\epsilon)/2\end{array}\right],
\end{multline}
and $F$ given by Eq.\ \eqref{F}. Apart from a factor $1/3$ in
$D_{22}$ and $E_{22}$ and a factor $2/3$ in $G_{22}$, which appear
as a direct consequence of the factors $1/3$ and $2/3$ in Eq.\
\eqref{decay}, this is exactly what is obtained by generalizing the
Gaussian treatment of spontaneous decay in Refs.\ \cite{hammerer},
\cite{madsen}, and \cite{sherson} to squeezing in a cavity.

Integrating the Ricatti equation numerically, we obtain the lower
curves in Fig.\ \ref{sqpmkavitet}, where $\Delta
p_{\textrm{at}}\equiv(\textrm{Var}(\hat{p}_{\textrm{at}}))^{1/2}$.
For the chosen parameters $\tilde{g}(\tau\ll
t\ll\eta(t\gg\tau)^{-1})\tau=2\cdot10^{-3}$,
$\kappa\tau=6\cdot10^{-3}$, and $\Phi\tau=3\cdot10^4$, and the
requirements of a dilute atomic gas, a high finesse cavity, and a
strong local oscillator are satisfied. The values
$\langle\hat{J}_x(0)\rangle/\hbar=5\cdot10^{11}\gg1$ and
$\langle\hat{a}_x(t\gg\tau)\rangle=4.6\cdot10^3\gg1$ justify the
classical treatment of these quantities, and $t=1\textrm{ ms}$
satisfies
$t\ll(4\eta_d\kappa_2\alpha(t\gg\tau)^2)^{-1}=2.7\cdot10^4\textrm{
s}$ and $t\ll\eta(t\gg\tau)^{-1}=0.13\textrm{ s}$. When atomic decay
is included, the uncertainty in $p_{\textrm{at}}$ does not decrease
indefinitely, but begins to rise at a certain point if the probing
is continued. For the given example, the minimum value of the
uncertainty is $(\Delta p_{\textrm{at}})_{\textrm{min}}=0.0233$.

For single pass squeezing \cite{madsen}
\begin{multline}
\left(\frac{d\textrm{Var}(\hat{p}_{\textrm{at}})}{dt}\right)_{\textrm{sp}}=
-2N_{\textrm{at}}\Phi\frac{g^4\tau^2}{\Delta^2}\eta_d\textrm{Var}
(\hat{p}_{\textrm{at}})^2e^{-\eta t}\\
-\frac{1}{3}\eta\textrm{Var}(\hat{p}_{\textrm{at}})+\frac{2}{3}\eta
e^{\eta t},
\end{multline}
and the result of an integration of this equation is shown in Fig.\
\ref{sqpmkavitet} for comparison. The squeezing is seen to occur on
a significantly slower time scale, and we note that $Q=5\cdot10^5$
for the chosen parameters. The attained minimum value of the
uncertainty $(\Delta p_{\textrm{at}})_{\textrm{min}}=0.121$ is also
significantly higher. This value is in accordance with the value
$0.118$ obtained from the approximate relation
\begin{equation}
(\Delta p_{\textrm{at}})_{\textrm{min}}=
\left(\frac{\eta\Delta^2}{3N_{\textrm{at}}\Phi
g^4\tau^2\eta_d}\right)^{1/4}
\end{equation}
derived in \cite{madsen} (we have included an additional factor of
$2/3$ to take the factors $1/3$ and $2/3$ in Eq.\ \eqref{decay} into
account). Since we found in Sec.\ \ref{III} that the main effect of
the cavity is to increase the squeezing rate by $Q$, and since it
follows from \eqref{eta} that $\eta$ is a factor
$4\kappa_1/((\kappa+\epsilon)^2\tau)$ larger for squeezing in a
cavity than for single-pass squeezing, we expect that $(\Delta
p_{\textrm{at}})_{\textrm{min}}$ is decreased by a factor
$((\kappa+\epsilon)^2\tau/(4\kappa_2))^{1/4}$ if the atoms are
enclosed in a cavity. This leads to the predicted value $(\Delta
p_{\textrm{at}})_{\textrm{min}}=0.0230$ for squeezing in a cavity,
which is close to the value observed in Fig.\ \ref{sqpmkavitet}.
Since $(\Delta p_{\textrm{at}})_{\textrm{min}}$ is proportional to
$N_{\textrm{at}}^{-1/4}$, we could also regard the squeezing
enhancement factor as a multiplicative factor on $N_{\textrm{at}}$,
and this opens the way to use the cavity to achieve measurement
induced squeezing of a smaller number of atoms. For
$N_{\textrm{at}}=10^{12}\cdot(\kappa+\epsilon)^2\tau/(4\kappa_2)=1.4\cdot10^9$
we thus find a minimum uncertainty of 0.121 after a probing time of
7 ms. We note that $\tilde{g}(t)$, $\eta(t)$, and $\epsilon$ are all
unchanged if $\Phi$, $N_{\textrm{at}}$, and $A$ are scaled by a
common factor, and we thus obtain the same result for $7\cdot10^6$
atoms if $\Phi=5\cdot10^{11}\textrm{ s}^{-1}$ and $A=10^{-6}\textrm{
m}^2$. A further decrease in $\Phi$ would, however, violate the
assumption of a classical $x$-polarized field and the approximation
below Eq.\ \eqref{Jx}.

\section{Conclusion}\label{V}

We have considered squeezing of one component of the collective spin
of an atomic ensemble achieved by performing homodyne measurements
on light, which has interacted with the atoms, and we have found
that the squeezing rate can be increased by a factor
$Q=16\kappa_1\kappa_2/(\kappa^4\tau^2)$ by placing the atoms inside
an optical cavity. For ensembles containing a large number of atoms
initially prepared in a coherent spin state, an efficient Gaussian
formalism is applicable, from which we have derived equations for
the time evolution of the covariance matrix describing the state of
the atomic spin and the cavity field, but we have also demonstrated
that analytical results for the state can be obtained even if the
state of the atomic spin is not Gaussian. Despite the stochastic
nature of the measurements, the variances of the components of the
atomic spin and the quadratures of the light field evolve
deterministically in the Gaussian approximation, and, in the
lossless case, the variance of the squeezed atomic spin component is
a monotonically decreasing function of time. According to the
Heisenberg uncertainty relation the variance of the conjugate atomic
spin variable has to increase, and the uncertainty product only
attains the smallest allowed value if all photons, transferred to
the mode with polarization orthogonal to the polarization of the
probe beam due to the interaction with the atoms, have left the
cavity and been detected at the considered time. Allowing the atoms
to decay spontaneously, we find that the minimum variance of the
squeezed spin component is obtained much faster and is approximately
reduced by a factor $((\kappa+\epsilon)^2\tau/(4\kappa_2))^{1/2}$
compared to the single-pass setup.

\end{document}